\documentclass[prl,twocolumn,showpacs,preprintnumbers,amsmath,amssymb]{revtex4}

\usepackage{graphicx}
\usepackage{dcolumn}
\usepackage{bm}

\def\bea{\begin{eqnarray}}
\def\eea{\end{eqnarray}}

\renewcommand{\d}{\partial}

\renewcommand{\geq}{\,{\geqslant}\,}
\renewcommand{\leq}{\,{\leqslant}\,}

\newcommand{\binner}[2]{%
  {\langle}\kern-4.15pt{\langle}#1{,}\,#2{\rangle}\kern-4.15pt{\rangle}}

\newcommand{\ffrac}[2]{\raisebox{.5pt}%
  {\footnotesize$\displaystyle\frac{#1}{#2}$}\kern1pt}

\def\cA{\mathcal{A}}

\def\cC{\mathcal{C}}

\def\cE{\mathcal{E}}

\def\cJ{\mathcal{J}}

\def\cL{\mathcal{L}}

\def\cQ{\mathcal{Q}}

\def\5{\bar }
\def\6{\partial }
\def\7{\hat }
\def\4{\tilde }

\def\Q#1#2{\frac{\partial #1}{\partial #2}}
\def\QS#1#2{\frac{\partial^S #1}{\partial #2}}
\def\varQ#1#2{\frac{\delta #1}{\delta #2}}

\def\eps{\epsilon}

\begin{document}


\title{Conserved charges and thermodynamics of the spinning G\"odel
  black hole}

\author{Glenn Barnich}

\author{Geoffrey Comp\`ere}
\affiliation{%
Physique Th\'eorique et Math\'ematique, \\ Universit\'e Libre de
    Bruxelles\\and\\ International Solvay Institutes\\ Campus
    Plaine C.P. 231, B-1050 Bruxelles, Belgium
}

\begin{abstract}
  We compute the mass, angular momenta and charge of the G\"odel-type
  rotating black hole solution to 5 dimensional minimal supergravity.
  A generalized Smarr formula is derived and the first law of
  thermodynamics is verified. The computation rests on a new approach
  to conserved charges in gauge theories that allows for their
  computation at finite radius.
\end{abstract}

\pacs{04.65.+e,04.70.-s,11.30.-j,12.10.-g}

\maketitle

Black hole solutions in supergravity theories have attracted a lot of
interest recently for two main reasons: on the one hand, higher
dimensional supersymmetric theories play a prominent role in the
effort of unifying gravity with the three microscopic forces; on the
other hand, black hole solutions are preferred laboratories to study
effects of quantum gravity.

Among the supersymmetric solutions of 5 dimensional minimal
supergravity \cite{Gauntlett:2002nw}, a maximally supersymmetric
analogue of the G\"odel universe \cite{Goedel:1949} has been found.
This solution can be lifted to 10 or 11 dimensions (see also
\cite{Tseytlin:1996as}) and has been intensively studied as a
background for string and M-theory, see
e.g.\cite{Boyda:2002ba,Harmark:2003ud}.

Black holes in G\"odel-type backgrounds have been proposed in
\cite{Herdeiro:2002ft,Gimon:2003ms,Herdeiro:2003un,%
  Brecher:2003wq,Behrndt:2004pn}.  Usually, given new black hole
solutions, the conserved charges are among the first properties to be
studied, see e.g.
\cite{Myers:1986un,Gauntlett:1998fz,Gibbons:2004ai}. Indeed, they are
needed in order to check whether these solutions satisfy the same
remarkable laws of thermodynamics as their four dimensional cousins
\cite{Bardeen:1973gs,Carter:1972}. The computation of the mass, angular
momenta and electric charge of the G\"odel black holes is an open
problem, mentioned explicitly in \cite{Gimon:2003ms} with partial
results obtained in \cite{Klemm:2004wq}, because the naive application of
traditional approaches fails. The aim of this Letter is to
solve this problem for the 5 dimensional spinning G\"odel-type black
hole \cite{Gimon:2003ms} and to derive both the generalized Smarr
formula and the first law.

\paragraph{Analytic expression for the charges.}
In odd space-time dimensions $n=2N+1$, the Einstein-Maxwell
Lagrangian with Chern-Simons term and cosmological constant reads
\begin{eqnarray}
L[g,A] && = \frac{\sqrt{-g}}{16\pi}[ R-2\Lambda - F_{\mu\nu}F^{\mu\nu}]
\nonumber\\
&&-\frac{2\lambda}{16\pi(N+1)\sqrt{3}}
\epsilon^{\gamma\alpha\beta\cdots\mu\nu}A_\gamma F_{\alpha\beta}\cdots
F_{\mu\nu} .\label{eq:1}
\end{eqnarray}
The bosonic part of $n=5$ minimal supergravity corresponds to
$\Lambda=0,\lambda=1$. The fields of the theory are collectively
denoted by $\phi^i\equiv(g_{\mu\nu}, A_\mu)$. Consider any fixed
background solution $\bar\phi^i$. The equivalence classes of conserved
$n-2$-forms of the linearized theory for the variables
$\varphi^i\equiv\phi^i-\bar\phi^i=(h_{\mu\nu}, a_\mu)$ can be shown
\cite{Barnich:2001jy} to be in one-to-one correspondence with
equivalence classes of field dependent gauge parameters
$\xi^\mu([\varphi],x),\epsilon([\varphi],x)$ satisfying
\begin{eqnarray}
\left\{\begin{array}{c}
\label{eq:2}
  \cL_\xi \bar g_{\mu\nu} = 0,\\ \cL_\xi \bar A_\mu + \partial_\mu
  \epsilon = 0,
\end{array}\right.
\end{eqnarray}
on-shell, i.e., when evaluated for solutions of the linearized theory.
Conserved $n-2$ forms are considered equivalent if they differ
on-shell by the exterior derivative of an $n-3$ form, while field
dependent gauge parameters are equivalent, if they agree on-shell. If
$n\geq 3$ and under reasonable assumptions on the background $\bar
g_{\mu\nu}$, the equivalence classes of solutions to the first
equation of \eqref{eq:2} are classified by the field independent
Killing vectors $\bar \xi^\mu(x)$ of the background $\bar g_{\mu\nu}$
\cite{Anderson:1996sc}. For the backgrounds that we
consider below, the second equation will then also be satisfied by
taking $\epsilon=0$. It is then straightforward to show that the
system \eqref{eq:2} admits only one more equivalence class of
solutions characterized by $\xi^\mu=0,\epsilon=c\in\mathbb{R}$.

One then computes the weakly vanishing Noether currents
\begin{eqnarray}
  \label{eq:4bis}
  S^\mu_{\xi,\epsilon} =
\varQ{L}{g_{\mu\nu}}(2\xi_\nu) + \varQ{L}{A_\mu}(A_\rho
\xi^\rho)+\varQ{L}{A_\mu} \epsilon,
\end{eqnarray}
associated with gauge transformations. For second order theories, the
$n-2$ forms $k_{\xi,c}[\varphi;\phi]=k_{\xi,c}^{[\mu\nu]}
(d^{n-2}x)_{\mu\nu}$ are defined through the formula
\begin{eqnarray}
k_{\xi,c}^{[\mu\nu]} &&  =  \frac{1}{2} \varphi^i
\Q{S^\mu_{\xi,c}}{\phi^i_\nu} \nonumber\\ &&
+[\frac{2}{3} \d_\lambda\varphi^i - \frac{1}{3} \varphi^i
\d_\lambda]\QS{S^\mu_{\xi,c}}{\phi^i_{\lambda \nu }} - (\mu\leftrightarrow
\nu )\label{eq:def_k},
\end{eqnarray}
with $(d^{n-p}x)_{\mu_1\dots\mu_p}= \frac 1{p!(n-p)!}\,
\epsilon_{\mu_1\dots\mu_n} dx^{\mu_{p+1}}\dots dx^{\mu_n}$ and
$\partial^S {\phi^j_{\mu \rho}}/\partial {\phi^i_{\lambda \nu}}
=\delta^j_i\delta^\lambda_{(\mu}\delta^\nu_{\rho)}$.  The forms
$k_{\xi,c}[\varphi;\bar\phi]$ are closed, $d
k_{\xi,c}[\varphi;\bar\phi]=0$, whenever $\bar\phi$ satisfies the
equations of motion, $\varphi$ the linearized equations of motion and
$(\xi,\epsilon)$ the system \eqref{eq:2}. For the solutions
$(\bar\xi,0)$, one can write $k_{\bar \xi,0} = k^{grav}_{\bar \xi} +
k^{em}_{\bar \xi} + \lambda k^{CS}_{\bar\xi}$.  The gravitational
contribution, which depends only on the metric and its deviations,
coincides with the Abbott-Deser expression \cite{Abbott:1982ff} and,
if one computes in the Hamiltonian framework, with the expression
derived in the Regge-Teitelboim approach \cite{Regge:1974zd}. Using
the Killing equation, it can be written in the Iyer-Wald form
\cite{Iyer:1994ys}:
\begin{eqnarray}
k^{grav}_{\bar \xi}[h;\bar g] &=& -\delta K^K_{\bar \xi}
-\bar\xi\cdot\Theta,\label{eq:6grav}
\end{eqnarray}
where
\begin{eqnarray}
K^K_{\bar\xi}&=&(d^{n-2}x)_{\mu\nu}
\frac{\sqrt{-g}}{16\pi}\Big(D^\mu\bar\xi^\nu-(\mu\leftrightarrow
\nu)\Big),\label{eq:7}
\end{eqnarray}
is the Komar $n-2$ form and
\begin{eqnarray}
\Theta&=&(d^{n-1}x)_\mu \frac{\sqrt{-\bar
  g}}{16\pi}\Big(\bar D_\sigma
h^{\mu\sigma}-\bar D^\mu h\Big).\label{eq:Theta}
\end{eqnarray}
Here and below, $\delta g_{\mu\nu}=h_{\mu\nu},\delta A_\mu=a_\mu$ and
after the variation, $(g,A)$ are replaced by $(\bar g,\bar A)$. We
also assume that the variation leaves $\bar \xi$ unchanged. After
dropping a $d$ exact form and using \eqref{eq:2}, the electromagnetic
contribution becomes
\begin{eqnarray}
k^{em}_{\bar \xi}[a;\bar A,\bar g] = -\delta Q^{em}_{\bar \xi,0}
-\bar \xi \cdot \Theta^{em}, \label{eq:6em}
\end{eqnarray}
where
\begin{eqnarray}
Q^{em}_{\bar\xi,c} &=&
(d^{n-2}x)_{\mu\nu}\frac{\sqrt{-g}}{4\pi} \left( F^{\mu\nu}(\bar\xi^\rho
  A_\rho+c)\right),\label{eq:8bis}\\
\Theta^{em}&=&(d^{n-1}x)_\mu \frac{\sqrt{-\bar
  g}}{4\pi}\Big(\bar F^{\alpha\mu} a_\alpha \Big).\label{eq:Thetaem}
\end{eqnarray}
The Chern-Simons term contributes as
\begin{equation}
k^{CS}_{\bar \xi}[a,\bar A] = - \frac{N(d^{n-2}x)_{\mu\nu}}
{4\sqrt{3}\pi}\eps^{\mu\nu\sigma\alpha\beta \cdots
\gamma\delta}a_\sigma \bar F_{\alpha\beta}\cdots \bar
F_{\gamma\delta}(\bar A_\rho \xi^\rho).\label{eq:6CS}
\end{equation}

For the solution $(0,1)$ of \eqref{eq:2} corresponding to the
conserved electric charge, we get, up to a $d$
exact term,
\begin{eqnarray}
k_{0,1}[a, h;\bar A,\bar g] = -\delta (Q^{em}_{0,1}+
\lambda
J),\label{eq:9}\\
J =\frac{(d^{n-2}x)_{\mu\nu}}{4\pi\sqrt{3}}\eps^{\mu\nu\sigma\alpha\beta\cdots
\gamma\delta}A_\sigma F_{\alpha\beta}\cdots
F_{\gamma\delta}.\label{eq:3bis}
\end{eqnarray}

Consider a path $\gamma$ in solution space joining the solution $\phi$
to the background $\bar \phi$. Let $\tilde\phi$ be a point on the path
and $\tilde\varphi$ a tangent vector at this point. Because
$k_{\xi,c}[\tilde\varphi,\tilde\phi]$ is closed if \eqref{eq:2} holds
with $\bar\phi$ replaced by $\tilde\phi$, it follows that
\begin{eqnarray}
  \label{eq:4}
  K_{\bar\xi,c}=\int_{\gamma} k_{\bar\xi,c}[d_V\phi;\phi],
\end{eqnarray}
with $d_V\phi$ a one-form in field space, is closed when integrated
along a path $\gamma$ in solution space as long as \eqref{eq:2} holds
for all solutions along the path \cite{Barnich:2003xg} (see also
\cite{Wald:1999wa}). Explicitly, if the path is parameterized by
$\phi^{(s)}$ for $s\in[0,1]$, we have
\begin{eqnarray}
  \label{eq:15}
  K_{\bar\xi,c}=\int^1_{0} k_{\bar\xi,c}[\varphi^{(s)};\phi^{(s)}],
\end{eqnarray}
with $\varphi^{(s)}=\frac{d}{ds}\phi^{(s)}$.
Whenever two $n-2$ dimensional closed hypersurfaces $S$ and
$S^\prime$ can be chosen as the only boundaries of an $n-1$
dimensional hypersurface $\Sigma$, the charges defined by
\begin{eqnarray}
Q_{\bar\xi,c}=\oint_S K_{\bar\xi,c}\label{eq:3}
\end{eqnarray}
do not depend on the hypersurfaces $S$ used for their evaluation.
Furthermore, the integrability conditions satisfied by
$k_{\bar\xi,c}[d_V\phi;\phi]$ imply, in the absence of topological
obstructions, that these charges do not depend on the path, but only on
the initial and the final solutions \cite{Barnich:2004uw}.

\paragraph{Mass, angular momenta and electric charge of G\"odel black
  holes.} We now assume $n=5,\Lambda=0,\lambda=1$. The G\"odel-type
solution \cite{Tseytlin:1996as,Gauntlett:2002nw} to the field
equations is given by
\begin{eqnarray}
\bar{ds}^2 &&= - (dt + j\,r^2\sigma_3)^2 + dr^2 +\nonumber\\&&+
\frac{r^2}{4}(d\theta^2+d\psi^2+d\phi^2+2\cos{\theta}d\psi d\phi),\\
\bar A &&= \frac{\sqrt{3}}{2}j\, r^2\sigma_3,\nonumber\label{GodelSol}
\end{eqnarray}
where the Euler angles $(\theta,\phi,\psi)$ belong to the intervals
$0\leq\theta \leq\pi$, $0\leq \phi \leq 2\pi$, $0\leq \psi < 4\pi$ and
where $\sigma_3 = d\phi+\cos{\theta}d\psi$. It is the reference
solution with respect to which we will measure the charges of the
black hole solutions of \cite{Gimon:2003ms} that we are interested
in. These latter solutions can be written
as
\begin{eqnarray}
&&ds^2 = \bar{ds}^2 + \frac{2m}{r^2}(dt-\frac{l}{2}\sigma_3)^2 \nonumber
-2m j^2r^2\sigma_3^2 \\ && \,\,\qquad + (k(r)-1)dr^2,
\qquad \quad A = \bar A,\label{eq:KerrGodel} \\
 \vspace*{-0.7cm} && k^{-1}(r) = 1 -\frac{2m}{r^2}+\frac{16j^2m^2}{r^2}
+ \frac{8jml}{r^2} +\frac{2ml^2}{r^4}.\nonumber
\end{eqnarray}
They reduce to the Schwarzschild-G\"odel black hole when $l =
0$, whereas the five dimensional Kerr black hole with equal rotation
parameters is recovered when $j = 0$.

For the charges defined through (\ref{eq:4}) and (\ref{eq:3}), we
choose to integrate over the surface $S$ defined by $t=cste=r$, while
the path $\gamma:(g^{(s)},A^{(s)})$ interpolating between the
background G\"odel-type universe $(\bar g,\bar A)$ and the black hole
$(g,A)$ is obtained by substituting $(m,l)$ by $(sm,sl)$ in
(\ref{eq:KerrGodel}), with $s\in [0,1]$.

Because $A^{(s)}_\mu=\bar A_\mu$ for all $s$, the mass
\begin{eqnarray}
\cE\equiv\oint_S K_{\frac{\partial}{\partial t},0}
\end{eqnarray}
of the black hole comes from the gravitational part only
\begin{eqnarray}
  \label{eq:6}
  \cE &=&- \left[ \oint_S K^K_{\Q{}{t}}\right]^{g}_{\bar g} - \int_0^1 ds\,
\oint_S \Q{}{t}\cdot\Theta[h^s;g^s]\nonumber\\
&=& \frac{3\pi}{4} m - 8\pi j^2\, m^2 -\pi j\ m\ l.
\end{eqnarray}
Unlike the 5 dimensional Kerr black hole
\cite{Myers:1986un,Gauntlett:1998fz}, the mass of which is
recovered for $j=0$, we also see that the rotation
parameter $l$ brings a new contribution to the mass with respect to
the Schwarzschild-G\"{o}del black hole.

Note that the integral over the path is really needed here in order to
obtain meaningful results, because the naive application of the
Abbott-Deser, Iyer-Wald or Regge-Teitelboim expressions gives as a
result
\begin{eqnarray}
  \label{eq:17}
  \cE^{naive}= \oint_S k_{\frac{\partial}{\partial t},0}
[g-\bar g,\bar g] = 8 \pi m^2 j^4 r^2 + O(1),
\end{eqnarray}
which, as pointed out in \cite{Klemm:2004wq}, diverges for large $r$.
A correct application consists in using these expressions to compare
the masses of infinitesimally close black holes, i.e., black holes
with $m+\delta m,l+\delta l$ as compared to black holes with $m,l$.
Indeed, $\oint_S k_{\frac{\partial}{\partial t},0} [\delta
g,g]=\delta\cE$, with $\cE$ given by the r.h.s of \eqref{eq:6}, which
is finite and $r$ independent as it should since
$dk_{\frac{\partial}{\partial t},0} [\delta g,g]=0$. Finite mass differences can
then be obtained by adding up the infinitesimal results.
This procedure is for instance also needed if one wishes to compute
in this way the masses of the conical deficit solutions  \cite{Deser:1983tn}
in asymptotically flat 2+1 dimensional
gravity.

Because our computation of the mass does not depend on the radius $r$
at which one computes, one can consider, if one so wishes, that one
computes inside the velocity of light surface. Similarly, if one uses
this method to compute the mass of de Sitter black holes, one can
compute inside the cosmological horizon, and problems of
interpretation, due to the fact that the Killing vector becomes
space-like, are avoided.

The expression for the angular momentum
\begin{eqnarray}
\cJ^\phi\equiv-\oint_S
K_{\frac{\partial}{\partial\phi},0}\label{eq:10}
\end{eqnarray}
reduces to
\begin{eqnarray}
\cJ^\phi &=& \left[ \oint K^K_{\Q{}{\phi}}\right]^g_{\bar g} + \left[ \oint
Q^{em}_{\Q{}{\phi},0} \right]^{A,g}_{\bar A,\bar g}
\nonumber\\&=&\frac{1}{2}\pi m
l -\pi j m l^2  -4 \pi j^2 m^2 l ,\label{eq:angul3}
\end{eqnarray}
while the angular momenta for the other 3 rotational Killing vectors
\cite{Gimon:2003ms} vanish.

The electric charge picks up a contribution from the Chern-Simons term
and is explicitly given by
\begin{eqnarray}
  \label{eq:11}
  \cQ\equiv-\oint_S K_{0,1}
=\left[ Q^{em}_{0,1}+{\lambda }J\right]^{g,\bar A}_{\bar
  g,\bar A} =
2\sqrt{3}\pi\, j m l.
\end{eqnarray}
In particular, it vanishes for the Schwarzschild-G\"odel black hole.

\paragraph{Generalized Smarr formula and first law.}
Consider a stationary black hole with Killing horizon determined by
$\xi_H=k+\Omega^H_a m^a$, where $k$ denotes the time-like Killing
vector, $\Omega^H_a$ the angular velocities of the horizon and $m^a$
the axial Killing vectors and let $\cE=\oint_S K_{k,0}$,
$\cJ^a=-\oint_S K_{m^a,0}$. As in \cite{Barnich:2004uw}, the
generalized Smarr relation then follows directly from the identity
\begin{eqnarray}
\oint_{S} K_{\xi_H,0} = \oint_{H} K_{\xi_H,0}\label{eq:14}
\end{eqnarray}
where $H$ is a $n-2$ dimensional surface on the horizon. Indeed, the
definition of $\xi_H$ and the charges imply
\begin{eqnarray}
  \label{eq:12}
  \cE - \Omega^H_a \cJ^a = \oint_H K_{\xi_H,0}.
\end{eqnarray}
Because $A^{(s)}_\mu =\bar A_\mu$, the r.h.s becomes
\begin{eqnarray} \label{eq:chargesmarr}
\oint_H K_{\xi_H,0}&=& -\left[\oint_H K^K_{\xi_H}\right]^{g}_{\bar
g} - \left[ \oint_H Q_{\xi_H,0}^{em} \right]_{\bar g,\bar A}^{g,\bar
  A} +
\oint_H\cC_{\xi_H;\gamma},\nonumber\\
\cC_{\xi_H;\gamma}&=& -\int_0^1 ds\, \xi_H \cdot
\Theta[h^{(s)};g^{(s)}]. \label{intcC}
\end{eqnarray}
Now, $-\oint_H K^K_{\xi_H}[g] = \frac{\kappa \cA}{8\pi}$, where
$\kappa$ is the surface gravity and $\cA$ the area of the horizon,
while $- \left[ \oint_H Q_{\xi_H,0}^{em}\right]_{\bar g,\bar
  A}^{g,\bar A} = \Phi_H \cQ$, where $\Phi_H = -(\xi_H \cdot A)$ is
the co-rotating electric potential, which is constant on the horizon
\cite{Carter:1972,Gauntlett:1998fz}. We thus get
\begin{eqnarray}
  \label{eq:13}
\hspace{-11pt}  \cE - \Omega^H_a \cJ^a \hspace{-4pt}&=&
\hspace{-4pt}\frac{\kappa \cA}{8\pi} +\Phi_H \cQ
 + \oint_H  K^K_{\xi_H}[\bar g] + \oint_H\cC_{\xi_H;\gamma} .
\end{eqnarray}

In order to apply this formula in the case of the black hole
(\ref{eq:KerrGodel}), we have to compute the remaining quantities.
The radius $r_H$ and the angular velocities $\Omega_\phi^H$ and
$\Omega_\psi^H$ are solutions of
\begin{eqnarray}
\left[ \Q{\xi^2}{\Omega^\phi}\right]_{r_H,\Omega^H_a} = 0,
\left[ \Q{\xi^2}{\Omega^\psi}\right]_{r_H,\Omega^H_a} = 0,
\left[\xi^2\right]_{r_H,\Omega^H_a} = 0.
\end{eqnarray}
Defining for convenience $\alpha  = (1-8j^2m)(1-8j^2m-8jl-2m^{-1}l^2)$ and
$\beta = 1-8j^2m-4r_H^2j^2+2ml^2r_H^{-4}$, we find
\begin{eqnarray*}
r_H^2 &=& m-4jml-8j^2m^2+ m \sqrt{\alpha}\\
\Omega^H_\phi& =& 4\frac{j +m lr_H^{-4} }{\beta},\qquad \Omega^H_\psi = 0.
\end{eqnarray*}
The electric potential is given by $\Phi_H = -(\xi_H \cdot \bar A)
= -\frac{\sqrt 3}{2}j r_H^2\Omega^\phi_H$. The area and surface
gravity of the horizon are
\begin{equation}
\cA = 2\pi^2r_H^3 \sqrt{\beta},\qquad \kappa = \frac{2m \sqrt{\alpha}
}{r^3_H\sqrt{\beta} }.
\end{equation}
For the G\"odel-Schwarzschild black hole, we recover the results of
\cite{Gimon:2003ms,Klemm:2004wq}:
\begin{eqnarray}
  \label{eq:16}
  r^2_H=2m(1-8j^2m),\ \cA =
  2\pi^2\sqrt{8m^3(1-8j^2m)^5},\nonumber\\
  \Omega^H_\phi=\frac{4j}{(1-8j^2m)^2},\
  \kappa = \frac{1}{\sqrt{2m(1-8j^2m)^3}}.\nonumber
\end{eqnarray}
Using
\begin{eqnarray}
\oint_H K^K_{\xi_H}[\bar g]= -\pi j^2 r_H^4 - \pi
j^3r_H^6\Omega_\phi^H,\\
\oint_H\cC_{\xi_H;\gamma} = \frac{\pi m}{4} -4\pi j^2m^2-\pi j m l+2\pi
j^2mr_H^2,
\end{eqnarray}
together with the explicit expressions for all the other quantities,
one can verify that the generalized Smarr formula~\eqref{eq:13}
reduces indeed to an identity.

We can also compare with the generalized Smarr formula derived
for asymptotically flat black holes in 5 dimensional
supergravity \cite{Gauntlett:1998fz}: for the G\"odel type black
hole (\ref{eq:KerrGodel}) we get
\begin{equation}
\frac{2}{3} \, \cE - \Omega_a \cJ^a -\frac{\kappa \cA}{8\pi} -
\frac{2}{3}\,\Phi_H \cQ =- \frac{2\pi}{3}jm(2jm+l).
\end{equation}
The right hand side, which vanishes when $j=0$, describes the breaking
of the Smarr formula for asymptotically flat black holes due to the
presence of the additional dimensionful parameter $j$. This is
somewhat reminiscent to what happens for Kerr-AdS black holes
\cite{Gibbons:2004ai}. In the latter case, different values of the the
cosmological constant $\Lambda$ describe different theories because
$\Lambda$ appears explicitly in the action. Even though this is not
the case for $j$, we have also taken $j$ here as a parameter
specifying the background because all charges have been computed with
respect to the G\"odel background.

As for Kerr-AdS black holes, the spinning G\"odel black hole satisfies
a standard form of the first law. Indeed, using the explicit
expressions for the quantities involved, one can now explicitly check
that the first law
\begin{equation}
\delta \cE = \Omega_a \delta \cJ^a + \Phi_H \delta \cQ +
\frac{\kappa}{8\pi}\delta \cA \label{first_law}
\end{equation}
holds. As pointed out in \cite{Gibbons:2004ai}, the validity of the
first law provides a strong support for our definitions of total
energy and angular momentum. Furthermore, in the limit of vanishing
$j$, we recover the usual expressions for 5 dimensional asymptotically
flat black holes.

\paragraph{Discussion.}

In the case of the non-rotating G\"odel black hole, $l=0=
\cJ^\phi=\cQ$, the parameterization $M^* = 2m-16j^2m^2$, $\beta^* =
\frac{2j}{1-8j^2m}$ suggested by the analysis of \cite{Gimon:2003xk}
allows one to write a non anomalously broken Smarr formula of the form
$\frac{2}{3} \, \cE^* =\frac{\kappa \cA}{8\pi}$, where
$\cE^*=\frac{3\pi}{8}M^*$, with $\kappa$ and $\cA$ unchanged. With
$\cE^*$ as the energy and $\beta^*$ the fixed parameter characterizing
the G\"odel background, the first law is however not satisfied. 

A way out, in the case $l=0$, is to consider the Killing vector
$k^\prime=(1+{\beta^*}^2M^*)^{-2/3}\frac{\partial}{\partial t}$. The
associated energy is $\cE^\prime\equiv\oint_S
K_{k^\prime}=\frac{3\pi}{8}M^*(1+{\beta^*}^2M^*)^{-2/3} $. The first
law now holds and in addition, with $\kappa^\prime$ defined with
respect to $k^\prime$, so does the non anomalously broken Smarr
formula $\frac{2}{3} \, \cE^\prime =\frac{\kappa^\prime
  \cA}{8\pi}$. Furthermore, it turns out that the
prefactor acts as an integrating factor and the first law is verified
for variations of both $M^*$ and $\beta^*$.  

\begin{acknowledgments}
  The authors are grateful to R.~Argurio, M.~Henneaux, J.~Zanelli,
  R.~Troncoso and C. ~Mart\'{\i}nez for useful discussions.  G.B.~and
  G.C.~are respectively Senior Research Associate and Research Fellow
  of the National Fund for Scientific Research, Belgium.  This work is
  supported in part by a ``P{\^o}le d'Attraction Interuniversitaire''
  (Belgium), by IISN-Belgium, convention 4.4505.86, by Proyectos
  FONDECYT 1970151 and 7960001 (Chile) and by the European Commission
  program MRTN-CT-2004-005104, in which the authors are associated to
  V.U.~Brussel.
\end{acknowledgments}


\end{document}